\documentclass[aps,prl,superscriptaddress,twocolumn,10pt,longbibliography]{revtex4-1}

\usepackage{amsmath}
\usepackage{graphicx}
\usepackage{bm}
\usepackage{color}

\date{\today}

\definecolor{darkblue}{rgb}{0.1,0.2,0.6}
\definecolor{darkred}{rgb}{0.8,0.1,0.2}
\usepackage[colorlinks,citecolor=darkblue,linkcolor=darkred,urlcolor=darkblue]
{hyperref}
\usepackage[all]{hypcap}

\usepackage{etoolbox}
\usepackage{mathtools}
\usepackage{amsmath}%
\usepackage{amsfonts}%
\usepackage{amssymb}
\usepackage[normalem]{ulem}

\begin{document}

\title{Slow many-body delocalization beyond one dimension}
\author{Elmer~V.~H.~Doggen}
\email[Corresponding author: ]{elmer.doggen@kit.edu}
\affiliation{\mbox{Institute for Quantum Materials and Technologies, Karlsruhe Institute of Technology, 76021 Karlsruhe, Germany}}
\author{Igor~V.~Gornyi}
\affiliation{\mbox{Institute for Quantum Materials and Technologies, Karlsruhe Institute of Technology, 76021 Karlsruhe, Germany}}
\affiliation{\mbox{Institut f\"ur Theorie der Kondensierten Materie, Karlsruhe Institute of Technology, 76128 Karlsruhe, Germany}}
\affiliation{A. F. Ioffe Physico-Technical Institute, 194021 St. Petersburg, Russia}
\author{\mbox{Alexander~D.~Mirlin}}
\affiliation{\mbox{Institute for Quantum Materials and Technologies, Karlsruhe Institute of Technology, 76021 Karlsruhe, Germany}}
\affiliation{\mbox{Institut f\"ur Theorie der Kondensierten Materie, Karlsruhe Institute of Technology, 76128 Karlsruhe, Germany}}
\affiliation{L.~D. Landau Institute for Theoretical Physics RAS, 119334 Moscow, Russia}
\affiliation{Petersburg Nuclear Physics Institute, 188300 St.~Petersburg, Russia}
\author{Dmitry~G.~Polyakov}
\affiliation{\mbox{Institute for Quantum Materials and Technologies, Karlsruhe Institute of Technology, 76021 Karlsruhe, Germany}}

\begin{abstract}
We study the delocalization dynamics of interacting disordered hard-core bosons for quasi-1D and 2D geometries, with system sizes and time scales comparable to state-of-the-art experiments.  The results are strikingly similar to the 1D case, with slow, subdiffusive dynamics featuring power-law decay.
From the freezing of this decay we infer the critical disorder $W_c(L, d)$ as a function of length $L$ and width $d$. In the quasi-1D case $W_c$ has a finite large-$L$ limit at fixed $d$, which increases strongly with $d$. In the 2D case $W_c(L,L)$ grows with $L$. The results are consistent with the avalanche picture of the many-body localization transition.
\end{abstract}

\maketitle

\emph{Introduction.---}
It was shown by Anderson \cite{Anderson1958a} that random disorder can localize non-interacting quantum particles that would classically exhibit diffusive behavior. A na\"ive expectation would be that, at finite temperature (or energy density), interactions between particles break this localization driven by quantum interference effects and establish ergodicity, so that the system satisfies the eigenstate thermalization hypothesis \cite{Polkovnikov2011a}. However, this is not always true. Due to the immense difficulty of systematically studying disordered many-body systems, a major breakthrough in the theoretical study of this problem had to wait until the mid-2000s \cite{Gornyi2005a, Basko2006a}. These analytical results, combined with early numerical works \cite{Oganesyan2007, Znidaric2008a, Pal2010a}, predicted a many-body localized (MBL) phase \cite{Nandkishore2015a, Altman2015a, Abanin2017a, Alet2018a} in which the system fails to thermalize. The breakdown of thermalization on experimental time scales was observed \cite{Schreiber2015a} in a system of ultracold atoms in a quasiperiodic potential.

In recent years, experiments have moved beyond one spatial dimension towards two-dimensional (2D) systems \cite{Choi2016a, Bordia2017a, Rubio-Abadal2019a, Chiaro2019a}, where the evidence for an MBL transition preceded by a slowing down of dynamics was also observed albeit at stronger disorder. From a numerical perspective, however, the 2D problem is extremely challenging. While exact diagonalization (ED) can reach around two dozen lattice sites in the one-dimensional (1D) case \cite{Luitz2015a, Pietracaprina2018a}, it is restricted to very small two-leg systems \cite{Baygan2015a, Zhao2017a,Wiater2018a}  and 1D systems coupled to a bath \cite{Hyatt2017a, Bonca2018a, Jansen2019a}, as well as extremely small ($4 \times 4$) 2D lattices \cite{Wiater2018a},  or requires constraints to reduce the Hilbert space \cite{Theveniaut2019a} to access physics beyond 1D.
The 2D case was studied using projected entangled pair states (PEPS), but this is restricted to very short times \cite{Kshetrimayum2019a} or yields mainly qualitative conclusions \cite{Kennes2018a}. Signatures of a MBL transition in two dimensions were found using a quantum circuit description \cite{Wahl2019a}. Two-dimensional models have also recently been studied using approximate perturbative approaches \cite{BarLev2016a, Thomson2018a, DeTomasi2019a}.

Here we show that using a 2D generalization of the time-dependent variational principle (TDVP) applied to matrix product states (MPS) allows for the systematic study of systems much larger than those accessible to ED, while reaching experimentally relevant, long time scales on the order of 100 hopping times. The method allows us to target both the strongly localized regime and the slow dynamics below the transition in a controlled way. The decay $\beta$ of the antiferromagnetic order with time $t$, quantified using the imbalance $\mathcal{I}(t)$, shows a power-law behavior $\mathcal{I}(t)\propto t^{-\beta}$ over this time window. Using extensive numerical simulations and exploiting the high accuracy of the TDVP compared to alternative MPS methods, we can determine $\beta$ with a statistical accuracy better than $10^{-2}$, permitting an accurate estimate for the critical disorder $W_c$ inferred from the vanishing of $\beta$ as a function of disorder strength $W$.

\begin{figure*}
 \includegraphics[width=\textwidth]{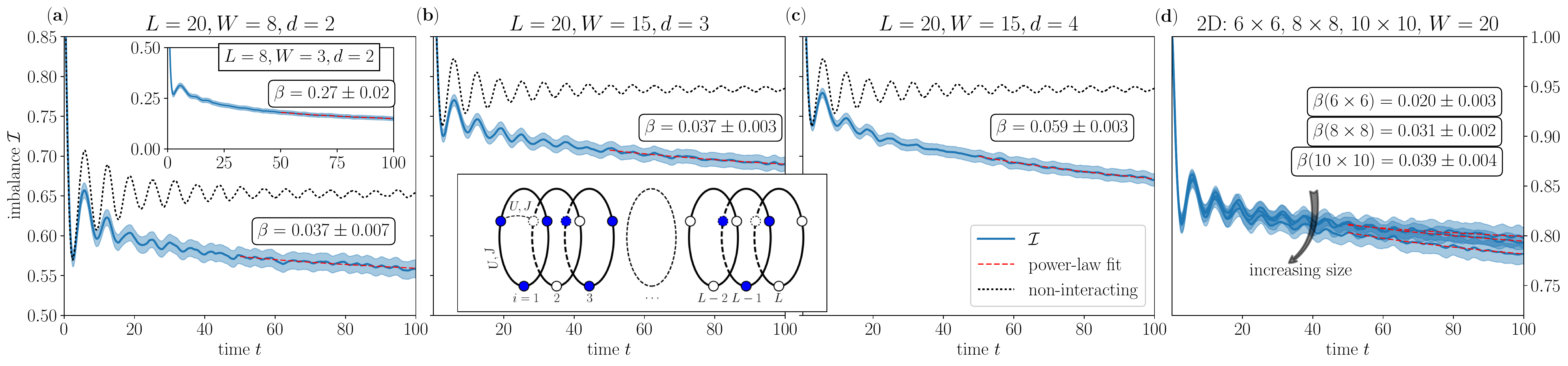}
 \caption{Time evolution of the columnar imbalance $\mathcal{I}$  averaged over disorder realizations for several representative cases on the ergodic side of the MBL transition for different choices of disorder strength $W$ and system dimensions $L$, $d$. The blue shaded region denotes a $2\sigma$-interval, the red dashed line is a power-law fit $\mathcal{I} \propto t^{-\beta}$, and the black dotted line is the exact result for the non-interacting case (10000 disorder realizations). Panel \textbf{(d)} shows the 2D case (note the differing $y$-axis range). \textbf{Inset of \textbf{(a)}}: Result using maximum bond dimension \cite{SupMat} for a weaker disorder and a smaller system, showing stronger decay. \textbf{Inset of \textbf{(b)}}: Depiction of the model \eqref{eq:ham} and initial condition for $d = 3$. Initially, particles occupy only odd-numbered columns $i$ (blue dots), while even $i$ are empty (white dots).}
 \label{fig:imbs}
\end{figure*}

In the 1D case, $\beta$ and similar power-law exponents derived from related transport quantities \cite{BarLev2015a, Agarwal2015a, Luitz2016a, Doggen2018a} are thought to be due to rare regions with anomalously strong disorder \cite{Agarwal2015a,Gopalakrishnan2016a} and their interplay with anomalously weakly disordered regions---``thermal spots''---driving the thermalization of the whole system through ``avalanches'' \cite{DeRoeck2017a,Thiery2017a, Dumitrescu2019a, Goremykina2019a}.
 The effect of these rare events is very sensitive to dimensionality  \cite{Gopalakrishnan2016a}, leading to the conclusion that such slow power-law dynamics should be absent in the long-time limit in two dimensions. On the other hand, the experimental data \cite{Bordia2017a} as well as the
 numerical solution of perturbative equations of motion \cite{BarLev2016a} indicate that subdiffusive behavior may survive in the 2D case.  Our approach sheds much needed numerical light on both experimental and analytical findings.

 Considering systems with dimensions $L \times d$ (length times width), with $L \ge d$, we determine the exponent $\beta$ and the critical disorder $W_c$ as functions of $L$ and $d$. We analyze $W_c(L,d)$ in two regimes: (i) quasi-1D systems with $L \gg d$ and (ii) 2D systems with $L=d$. It is crucial that, even though the system delocalizes at any $W$ in the thermodynamic limit, $W_c(\infty,\infty) = \infty$, one can define the transition point $W_c(L,d)$ for $L \gg 1$ and arbitrary $d$. $W_c(L,d)$ carries information about the mechanism of many-body delocalization. For $W < W_c(L,d)$ we define the size-dependent power-law exponent $\beta(L,d)$. While $\beta$ can in principle also depend on time scale (as the theory \cite{Gopalakrishnan2016a} predicts), we do not observe any such drift within the time range explored. Importantly, our characteristic times and system sizes are of the same order as in experiment, thus allowing a direct comparison.

\emph{Model and method.---}
We consider hard-core bosons with nearest-neighbor interactions on a 2D square lattice with length $L$ and width $d$ and spatial indices $i = 1, \ldots,L$ and $j = 1,\ldots,d$. The Hamiltonian is given by:
\begin{equation}
 \mathcal{H} =  \sum_{\langle ij;i'j' \rangle} \left[ -\frac{J}{2} (b_{ij}^\dagger b_{i'j'} + \mathrm{h.c.})  + U\hat{n}_{ij} \hat{n}_{i'j'} \right] + \sum_{ij} \epsilon_{ij} \hat{n}_{ij}.
  \label{eq:ham}
\end{equation}
Here $b_{ij}^{\dagger}$ creates a boson on a site $(i,j)$,
$\hat{n}_{ij} \equiv b_{ij}^\dagger b_{ij}$, and the summation over $\langle ij;i'j' \rangle$ couples neighboring sites, with periodic boundary conditions in the $j$-direction and open boundary conditions in the $i$-direction, see Fig.~\ref{fig:imbs}. On-site potentials
$\epsilon_{ij}$  are independent random variables taken from a uniform distribution $[-W, W]$. The occupation of each site is restricted to $n_{ij} \leq 1$.
We choose $J = U = 1$.

For our numerical simulations, we employ the time-dependent variational principle (TDVP) \cite{Haegeman2016a}. It projects the time evolution back onto the variational manifold of the matrix product state (MPS):
\begin{equation}
 \partial_t |\psi \rangle  = -i \mathcal{P}_\mathrm{MPS}\mathcal{H}|\psi\rangle. \label{eq:tdvp}
\end{equation}
The TDVP has been shown to be an efficient all-round MPS-based method \cite{Paeckel2019a} and works especially well for disordered systems \cite{Kloss2018a, Doggen2018a, Doggen2019a, Chanda2019a}.
A key advantage is that time evolution after the projection $\mathcal{P}_\mathrm{MPS}$ conserves the wave function norm and energy \cite{Haegeman2016a}.
This is in contrast to alternative MPS-based methods for time evolution \cite{Vidal2003a,Daley2004a,Zaletel2015a}. We use both a two-site implementation and a hybrid \cite{Goto2019a} one-site/two-site implementation of TDVP, see \cite{SupMat} for details.

\emph{Results.---}
We follow the quench dynamics  for a system prepared  in an initial state with occupied odd-numbered columns (see Fig.~\ref{fig:imbs}). Similar to the experiment \cite{Bordia2017a,Rubio-Abadal2019a}, we track the ``columnar imbalance'' $\mathcal{I} = (n_\mathrm{odd} - n_\mathrm{even})/(n_\mathrm{odd} + n_\mathrm{even})$, where $n_\mathrm{odd(even)}$ is the total particle density at odd (even) columns. For an ergodic system in the thermodynamic limit, the time-average of $\mathcal{I}(t)$ will approach zero as $t \rightarrow \infty$, since in that case the memory of the initial state is lost. We assess the decay of $\mathcal{I}$ by fitting to a power law $\mathcal{I}(t) \propto t^{-\beta}$ over the time window $t \in [50, 100]$. As we will see, the decay closely follows a power-law behavior in this time interval; we will discuss possible deviations below.
For each $(L,d)$, we determine the dependence of the exponent $\beta$ on disorder $W$. The critical disorder $W_c(L,d)$ is estimated as the value of $W$ at which $\beta$ vanishes \cite{SupMat}.

For $d=1$ and $L = 16$ this procedure yields a value for the critical disorder $W_c(16,1) \approx 4$ \cite{Doggen2018a} that is in good agreement with numerical results obtained from ED \cite{Luitz2015a, Luitz2016a} on systems of a comparable size. At the same time, the value found by this approach in Ref.~\cite{Doggen2018a} is substantially higher, $W_c(100,1) \approx 5.5$. While this demonstrates that finite-size effects are rather pronounced already in 1D \cite{Bera2017a}, we will see that they become qualitatively more important in quasi-1D and 2D.

\begin{figure}
 \includegraphics[width=\columnwidth]{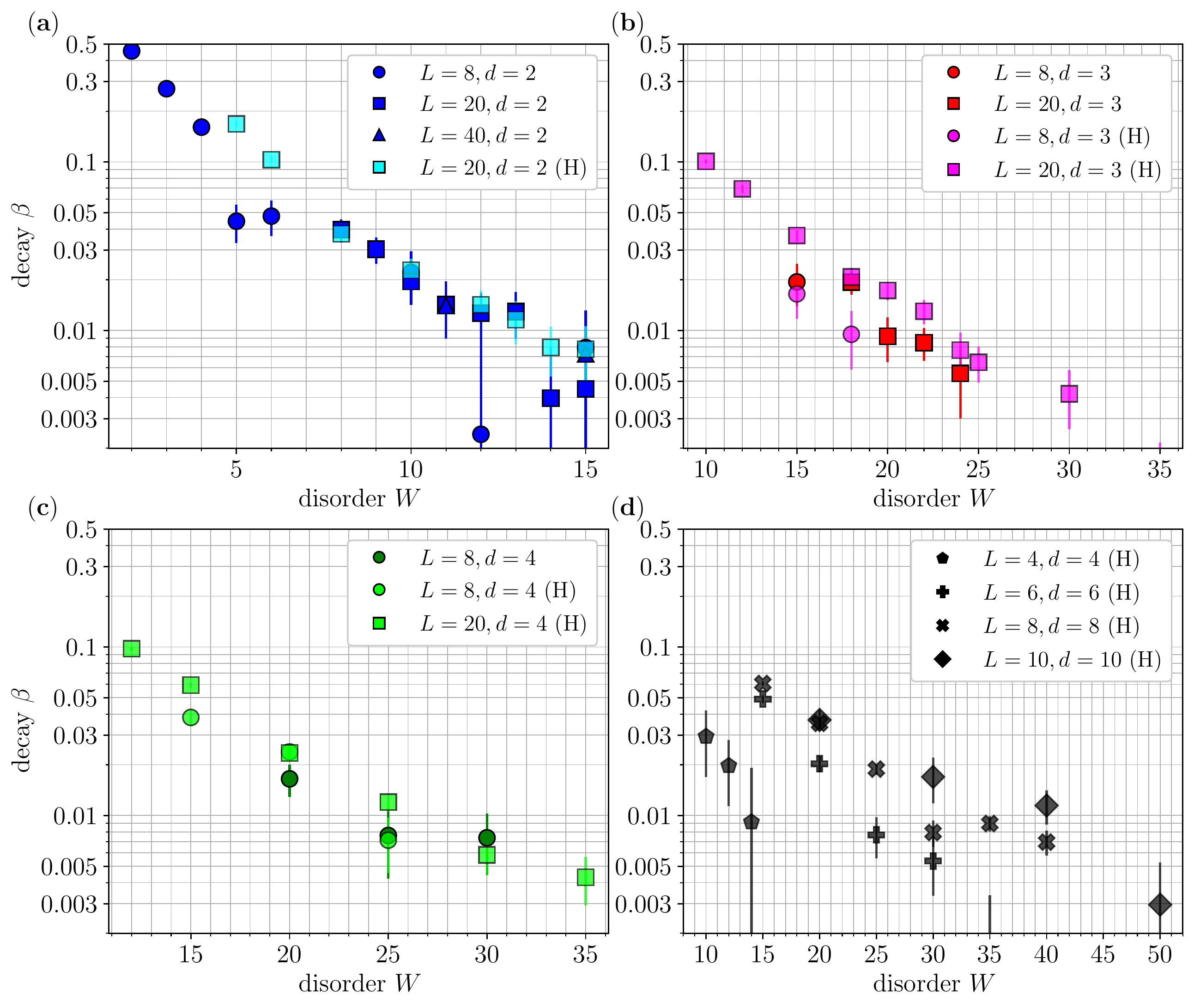}
 \caption{Power-law exponent $\beta$ characterizing decay of the disorder-averaged columnar imbalance $\mathcal{I} \propto t^{-\beta}$ in the time window $t \in [50,100]$, as a function of the disorder strength $W$. Panels \textbf{(a)-(c)}:  quasi-1D systems with $d=2$, 3, and 4 respectively for various $L$. Panel \textbf{(d)}: 2D systems with $L=d$.
 Error bars are $1\sigma$-intervals. Data labeled with ``(H)'' is obtained using the hybrid algorithm, the two-site algorithm is used otherwise. Independent disorder realizations are used for each data point. Data for $4\times4$ and $8\times2$ uses the maximum bond dimension \cite{SupMat}. Note the differing $x$-axis ranges.}
 \label{fig:beta}
\end{figure}

Several representative cases of the imbalance dynamics $\mathcal{I}(t)$ are shown in Fig.~\ref{fig:imbs}.  We find that with increasing transverse dimension $d$, the disorder required to freeze the dynamics is strongly increased compared to a 1D chain ($d=1$). At the same time, the imbalance decay remains well approximated by a power law  in the slowly thermalizing regime on the delocalized side of the transition. We show the obtained power-law exponent $\beta$ as a function of $W$ in Fig.~\ref{fig:beta} for quasi-1D systems with $d=2$, 3, and 4, and for 2D systems with values of $L=d$ up to 10. In the 2D case, we show $\mathcal{I}(t)$ curves for different $L$ on one plot in order to visualize an increase of $\beta$ (i.e., enhancement of delocalization) with the system size.

\begin{figure}
 \includegraphics[width=\columnwidth]{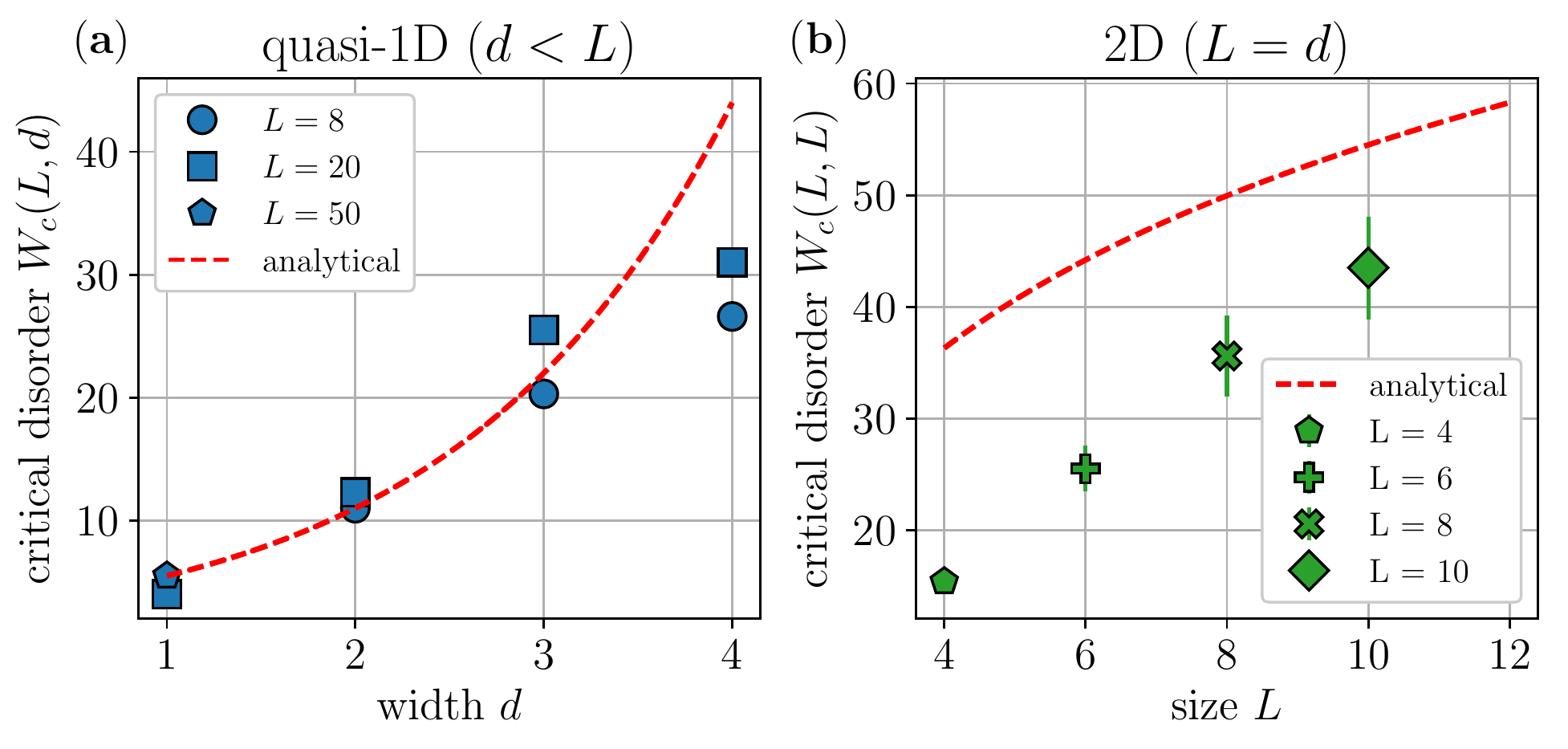}
 \caption{Estimated critical disorder $W_c(L,d)$ as a function of system size in the \textbf{(a)} quasi-1D and \textbf{(b)} 2D cases. Data for $d=1$ in the left panel are from Ref.~\cite{Doggen2018a}. For $d=1$, the value for $L=100$ is the same as for $L=50$. For $d=2$, the value for $L=40$ is the same as for $L=20$. (These additional data points are not shown for the sake of clarity.)
 Lines: analytical predictions for \textbf{(a)} $W_c(\infty,d)$ with $d\gg 1$ and \textbf{(b)} $W_c(L,L)$ with $L\gg 1$, see Eqs.~(\ref{eq:2dWc}) and (\ref{eq:q1dWc}).  \label{fig:Wc} }
\end{figure}

Obtained results for power-law exponents $\beta(L,d; W)$ as functions of disorder $W$ are shown in  Fig.~\ref{fig:beta}. Resulting values of the critical disorder $W_c(L,d)$ are presented in  Fig.~\ref{fig:Wc}. In panel \textbf{(a)}, we show $W_c(L,d)$ for quasi-1D systems, where we also included data for strictly 1D systems, $d=1$, from Ref.~\cite{Doggen2018a}.
In that paper, system sizes $L=20$, 50, and 100 were studied. The results for $L=50$ and 100 were nearly indistinguishable, which means that the critical disorder $W_c(L,1)$ saturates  at such values of $L$. Thus, the obtained value $W_c(50,1) \approx W_c(100,1) \approx 5.5$ serves as a good estimate for the thermodynamic-limit critical disorder $W_c(\infty,1)$.

For $d=2$, we studied systems of length $L=8$, 20, and 40. The value of $W_c$ for $L=20$ is slightly higher (by approximately 10\%) than that for $L=8$. The values for $L=20$ and $L=40$ are almost identical. Thus, we conclude that $W_c(L,2)$ has reached saturation at these values of $L$, so that our result $W_c(20,2) \approx 12$ is also a good estimate for $W_c(\infty,2)$.

For $d=3$ and 4 we obtain $W_c(20,3) \approx 26$ and $W_c(20,4) \approx 31$. These values
are higher by $20 - 30 \%$ than $L=8$ values. Since we have not studied systems of length larger than $L=20$, we cannot verify explicitly whether $W_c$ has reached saturation at this length, which means that the $L\to\infty$ values can be still somewhat higher. At the same time, analytical arguments (see below) indicate that for $d \le 4$ the length $L$ at which the thermodynamic limit is essentially reached should not be too large. We thus argue that our above results for $W_c(20,3)$ and $W_c(20,4)$ serves not only as lower bounds  for $W_c(\infty,3)$ and $W_c(\infty,4)$, respectively, but also as reasonable estimates for them. Comparing the obtained estimates for $W_c(\infty,d)$ with $d=1$, 2, 3, and, 4, we observe a fast increase of critical disorder with $d$.

Fig.~\ref{fig:beta}\textbf{(d)} displays results for 2D systems ($L=d$). We observe a clear increase of $W_c(L,L)$ with $L$, with respective estimates of $W_c \approx \{15, 26, 36, 44\}$ for $L = \{4, 6, 8, 10\}$. Below, we will compare our numerical findings for quasi-1D and 2D systems with analytical expectations.

\emph{Avalanches.---}
Recently, a phenomenological description for the many-body delocalization was proposed \cite{DeRoeck2017a,Thiery2017a}, which relies on the ``avalanche'' instability, i.e., proliferation of an initial effectively thermal seed which grows until it ``swallows'' the whole system  for $W < W_c$. Here we summarize the implications of this idea for our model, with additional details provided in \cite{SupMat}.

\begin{figure}
 \includegraphics[width=\columnwidth]{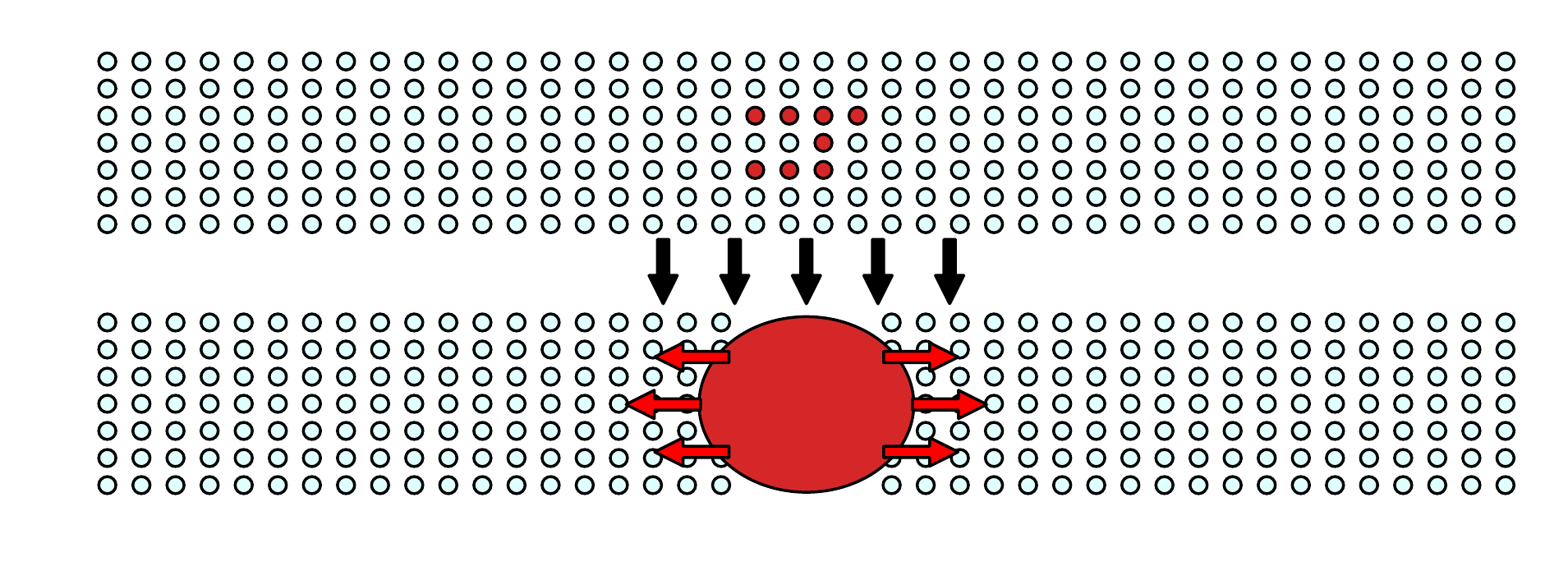}
 \caption{Avalanche instability in a quasi-1D geometry for $L \ll  L_*(d)$.  For $W < W_c(L,d)$, an initial seed, a region of unusually weak disorder (top, red dots) thermalizes its surroundings and shows a 2D growth. Once the ergodic spot thus created reaches the size $d$, it grows along the $L$ direction (bottom), eventually thermalizing the whole system.}
 \label{fig:avalanche}
\end{figure}

The effect of avalanches on a 2D system ($L=d$ in our notations) was considered in Ref.~\cite{Gopalakrishnan2019a}. For a 2D system of size $L \gg 1$,  the critical disorder is given by
\begin{equation}
W_c(L,L)\sim \exp(c_1\ln^{1/3}L^2),
\label{eq:2dWc}
\end{equation}
up to subleading factors. An estimate presented in \cite{SupMat} yields $c_1 = 2 (\ln 2)^{2/3} \approx 1.57$.
While $W_c(L,L)$ grows without bound with increasing $L$, the transition point is well defined in a system of given (large) size $L$.
Generalization of the analysis of Refs.~\cite{Thiery2017a,Gopalakrishnan2019a} on a strip of width $d \gg 1$ and length $L > d$ yields \cite{SupMat}
\begin{equation}
W_c(L,d) \sim \begin{cases} \exp[c_1\ln^{1/3}(Ld)], &\quad d<L<L_*(d),\\
2^d, &\quad L>L_*(d),
\end{cases}
\label{eq:q1dWc}
\end{equation}
where the crossover length $L_*(d)$ reads
\begin{equation}
L_*(d) \sim d^{-1} \exp[(d \ln 2/c_1)^3].
\label{eq:Lstar}
\end{equation}
For $L \ll L_*(d)$ the avalanche is a two-step process: first a 2D growth of the seed until its size reaches $d$, then 1D growth. The bottleneck for thermalization in this case is the first (2D) stage, as schematically depicted in Fig.~\ref{fig:avalanche}, which is why the first line of Eq.~(\ref{eq:q1dWc}) is very similar to Eq.~(\ref{eq:2dWc}). On the other hand, for $L \gg L_*(d)$, a necessary rare ergodic spot of size $d$ can be found. In this case, the  bottleneck is an effectively 1D propagation of an avalanche with a thickness $d$.

As seen from Eq.~(\ref{eq:q1dWc}), for any given $d$ the critical disorder saturates in the limit $L\to\infty$.  At the same time, to reach the limiting value $W_c(\infty,d)$, one needs systems longer than $L_*(d)$.  For $d \le 3$  Eq.~(\ref{eq:Lstar}) yields $L_*(d)$ of order unity, so that the condition reduces to $L \gg 1$; for $d=4$ we find $L_*(4) \approx 60$.  Indeed, as we know from numerics  of Ref.~\cite{Doggen2018a} and of the present work, $L=50$ is sufficient to reach saturation for $d=1$, and $L=20$ is sufficient for $d=2$. It is expected, in view of the smallness of $L_*(d)$,  that similar values of $L$ should be sufficient for $d=3$ and $d=4$.
At the same time, for $d \ge 5$ the length $L_*(d)$ as given by Eq.~(\ref{eq:Lstar}) increases dramatically, implying that such systems should be in the 2D regime [first line of Eq.~(\ref{eq:q1dWc})] for any  $L$ that is realistic for numerical simulations or experiment.

In Fig.~\ref{fig:Wc}, we show by lines analytical results for $W_c(\infty,d)$, second line of  Eq.~(\ref{eq:q1dWc}) and $W_c(L,L)$, Eq.~(\ref{eq:2dWc}). Although these formulas are actually derived in the asymptotic limit of large $d$ (respectively, large $L$), we observe that they compare well with data already for relatively small values of these parameters.

It is worth noting that our numerical estimates of $W_c(\infty,d)$ for $d=2$ and $3$ are considerably higher than  those obtained in Refs.~\cite{Wiater2018a,Baygan2015a} by ED of small systems \cite{SupMat}. This is in full consistency with a drift of $W_c(L,d)$ with $L$; as a result, small-system data substantially underestimate $W_c(\infty,d)$.

\emph{Form of the long-time dynamics.---}
On the delocalized side of the transition, the Griffiths picture of delocalization \cite{Agarwal2015a, Gopalakrishnan2016a, Luitz2017a, Gopalakrishnan2019b, Lenarcic2019a}, which also applies in the avalanche scenario, predicts the following form of imbalance decay $\mathcal{I}(t)$ in $D$ dimensions:
\begin{equation}
 \mathcal{I}(t) \propto \exp\Big(-\gamma \ln^D t\Big), \label{eq:2d}
\end{equation}
where $\gamma$ is a constant. This leads to a power-law decay with the exponent $\beta = \gamma$ in 1D and to a somewhat faster decay $\mathcal{I}(t) \propto t^{-\gamma \ln t}$ in 2D. This form is almost indistinguishable from a pure power law for the time scale we consider here, and both forms provide essentially equally good fits \cite{SupMat}. Furthermore, the exponent $\gamma$ is rather small for all values of disorder considered here for the 2D case (see Fig.~\ref{fig:beta}), so that any term of the type \eqref{eq:2d} would in fact dominate over hydrodynamic power-law long-time tails up to any experimentally relevant time scale (cf.~Ref.~\cite{Bordia2017a}). Hence,
even though Griffiths effects are sub-leading in $D > 1$, they describe the physics in a rather broad range of disorder on the ergodic side of the transition up to extremely long time scales.

\emph{Summary and outlook.---} We have investigated, by means of the TDVP, the delocalization dynamics of interacting hard-core bosons  in quasi-1D and 2D disordered lattices for system sizes  comparable to state-of-the-art experiments.
We provide numerical results for the density imbalance dynamics for system sizes much larger than those accessible to ED.
Our results demonstrate subdiffusive, approximately power-law decay of imbalance on experimentally relevant time scales on the ergodic side of the MBL transition beyond 1D geometry. We have further determined with a high precision the power-law exponent $\beta$ as a function of disorder $W$ and the system width $d$ and length $L$ in the 2D ($L=d$) and quasi-1D ($L > d$) cases. Using these data, we have found the critical disorder $W_c(L,d)$ at which the dynamics freezes. Our findings support the expected saturation of $W_c(L,d)$ in the limit $L \to \infty$ at fixed $d$. Further, our results are consistent with analytical predictions (based on the avalanche picture of the MBL transition \cite{DeRoeck2017a}) for unbounded growth of $W_c(\infty,d)$ with $d$ in the quasi-1D case and of $W_c(L,L)$ with $L$ in the 2D case.

Our work opens important avenues for investigation of slow quantum dynamics and MBL in a variety of quasi-1D and 2D systems.
One direction of interest is the study of soft-core disordered Bose-Hubbard models \cite{Choi2016a, Sierant2018a, Chiaro2019a, Orell2019a, Hopjan2020a}. The Fermi-Hubbard model has substantially more complicated localization properties \cite{Prelovsek2016a}, exhibiting spin-charge separation in the delocalization dynamics (unless there is additional disorder in the spin degree of freedom \cite{LeipnerJohns2019a}). Furthermore, our approach can be straightforwardly extended to long-range interacting systems, which is of particular interest in context of experiments on trapped ions \cite{Smith2016a}.

\acknowledgments{
\emph{Acknowledgments}.--- We thank S.~Gopalakrishnan, M.~Heyl, S.~Rex, T.~Wahl, and J.~Zakrzewski for useful discussions.
MPS simulations have been carried out using the TeNPy library (versions \texttt{v0.4.0.dev0+0c4ad94} and \texttt{0.6.1.dev18+037ad76}) \cite{Tenpy} and the Open Source Matrix Product States library (version \texttt{v.2.1.r133}) \cite{Wall2012a,Jaschke2018a}, implementing the hybrid and two-site TDVP algorithms, respectively \cite{Haegeman2016a}.
Figures were generated using Matplotlib \cite{Matplotlib}. Simulations for the non-interacting model were performed using the NumPy \cite{NumPy} implementation of Lapack \cite{Lapack1999}. The authors acknowledge support by the State of Baden-W\"urttemberg through bwHPC.
}


\bibliography{ref}

\end{document}


\title{Supplementary Material to ``Slow many-body delocalization beyond one dimension''}

\maketitle

\renewcommand{\theequation}{S\arabic{equation}}
\renewcommand{\thefigure}{S\arabic{figure}}
\setcounter{equation}{0}
\setcounter{figure}{0}

In this Supplementary Material, we outline in detail the ``avalanche'' description of the MBL transition, the protocol of our numerical implementation, the analysis of data, and comparison with numerical estimates of $W_c$ in quasi-1D systems.

\section{Avalanche instability}

Here we discuss the ``avalanche''  theory of the MBL transition as applied  to the quasi-1D and 2D cases of the Hamiltonian considered in the main text. The reader is referred to Refs.~\cite{DeRoeck2017a, Thiery2017a, Gopalakrishnan2019a,Gopalakrishnan2019b} for previous works on the avalanche theory.

We consider   the model (1) of the main text
with $J=U=1$ (hopping and interaction) and $W\gg 1$ (disorder) on a strip of length $L$ and width $d$.
Assume that the system contains a rare region with anomalously weak disorder.
In the absence of couplings to the rest of the system, this region would be ergodic.
The exact many-body states in this ergodic seed are described by random matrix theory.
The rest of the system, with the ergodic seed excluded, is assumed to be many-body localized.
The Hamiltonian of the localized part of the system is then described in terms of mutually commuting
operators, each representing a  local integral of motion (LIOM), a.k.a. $l$-bit.

At the heart of the avalanche theory is the observation that a single ergodic spot (originating from a rare region where disorder is anomalously weak) can delocalize the otherwise many-body localized system but successively absorbing sites around it---the process that was termed an ``avalanche''. This happens under certain conditions on the size of the rare spot and the typical localization length (and thus strength of disorder) in the surrounding MBL system. The probability to find a rare spot of a required size depends on system dimensions. As a result, one can determine the dependence of the critical disorder $W_c$ (below which an avalanche develops) on $L$ and $d$. Below we outline this analysis first for the 2D case ($L=d$) and then for the quasi-1D geometry ($L \gg d$).

\subsection{Avalanche in 2D}

For an avalanche \cite{DeRoeck2017a} to start, the ergodic seed should exceed the critical size
which can be estimated following Ref.~\cite{DeRoeck2017b} [see the discussion around Eq. (3.4) there].
We begin with the 2D case ($L=d$), assuming a square lattice.
From the condition $e^{-1/\xi}\sim 1/W$, we estimate the typical localization length
\begin{equation}
\xi\simeq 1/\ln W.
\label{loc-length}
\end{equation}
For a 2D lattice this result corresponds to the shortest path (``forward approximation'') between the two points along the links of the lattice.

Consider an ergodic seed of radius $R_s$ (counted along the links of the square lattice from the central point). Such a ``lattice sphere'' is, in fact, a square rotated by $\pi/4$ with respect to the original square lattice, with the diagonal equal to $2R_s$. The volume of this ``sphere'' is $2 R_s^2$.
The ratio of the matrix element to the relevant (many-body) level spacing for the transition of a spin ($l$-bit) located at distance $r$ (again, counted along the links of the lattice) from the seed scales as
\begin{equation}
\frac{\gamma^r N_s^{-1/2}}{N_s^{-1}}\gtrsim 1,
\label{ratio}
\end{equation}
where $\gamma\sim e^{-1/\xi}\sim 1/W$ and $N_s$ is the dimension of the space of many-body states inside the 2D seed. The factor $N_s^{-1/2}$ in the matrix element accounts for the correct scaling of the local operators in the ergodic seed. The condition (\ref{ratio}) defines the size of the ``buffer zone'' around the seed,
\begin{equation}
r(R_s)\sim R_s^2/|\ln \gamma| \sim  R_s^2/\ln W,
\label{rRs}
\end{equation}
where the resonance condition is satisfied for all the spins.

In the limit of large $R_s$, when $r(R_s)\gtrsim R_s$, the ergodic seed grows indefinitely by absorbing the buffer spins: an avalanche occurs. This defines the critical size of an ergodic seed:
\begin{equation}
R_{c}\sim \ln W.
\label{Rc-est}
\end{equation}
Once an ergodic seed is found, the whole 2D system is thermalized by the avalanche.
The number of spins in the critical seed is
\begin{equation}
n_c\sim \ln^2 W.
\label{nc-est}
\end{equation}

In order to fix the coefficient in Eq.~(\ref{nc-est}),
we proceed following Ref. \cite{DeRoeck2017a}. We include all the hybridized spins into the dimension of the space of ergodic many-body states:
\begin{equation}
N_s=2^{2 (R_s+r)^2}.
\label{NsRsr}
\end{equation}
The condition for the avalanche to proceed for arbitrary $r$ then reads
\begin{equation}
F(r)=\frac{\ln2}{\ln W}(R_s+r)^2-r>0.
\end{equation}
For small enough $R_s$, this condition is not fulfilled at not too large $r$ and hence the avalanche does not develop. Then $R_c$ is found from the requirement that the minimum of $F(r)$ corresponds to $F>0$,
yielding
\begin{equation}
R_{c}\simeq \frac{\ln W}{4\ln 2}
\label{Rc}
\end{equation}
and
\begin{equation}
n_c\simeq c \ln^2 W,
\label{ns}
\end{equation}
with
\begin{equation}
 c=\frac{1}{8 \ln^2 2}.
 \label{cc}
\end{equation}

The probability of finding a single critical seed in a 2D system of linear size $L$ is
\begin{equation}
P(L,L;W)\sim L^2 \left(\frac{1}{W}\right)^{n_c}
\label{Prob}
\end{equation}
(each of the $n_s \gg 1$ spins of the seed should have an energy within the window $\sim 1$).
Using Eq. (\ref{ns}), we get
\begin{equation}
P(L,L;W)\sim L^2 W^{- c \ln^2 W}.
\label{PLW}
\end{equation}
Equating this probability to $1/2$, we estimate the critical disorder strength for $D=2$, Ref. \cite{Gopalakrishnan2019a}:
\begin{equation}
\ln W_c(L,L)\simeq c_1 \ln^{1/3} L^2,
\label{WLL}
\end{equation}
where $c_1= c^{-1/3}$.
The width of the transition can be estimated \cite{Gopalakrishnan2019a} by comparing the values of $W$ for which $P=1/4$
and $P=3/4$:
\begin{equation}
\frac{\delta W(L,L)}{W_c(L,L)}\sim \ln^{-2/3}L.
\label{deltaWLL}
\end{equation}
It is seen that with increasing $L$, the critical disorder strength grows to infinity, while the transition
sharpens.

\subsection{Quasi-1D avalanche}

Let us now turn to the quasi-one-dimensional case, $L\gg d\gg 1$.
We first consider the case when the critical size of the 2D seed (\ref{Rc}) is smaller than the width
of the strip:
\begin{equation}
d \gg \ln W.
\label{2Dseed}
\end{equation}
In this case, the avalanche consists of the two stages: first the critical
seed explodes as in the 2D case, but after reaching the boundary of the strip the avalanche continues essentially in the 1D manner (the only difference compared to the true 1D case is that the ergodic spot now has $d$ neighbors).  The probability of finding a single critical 2D seed in the sample of area $L\times d$
is
\begin{equation}
P(L,d,W)\sim L d\, W^{-c \ln^2 W}.
\label{PLdW}
\end{equation}
Thus, the necessary condition for starting an avalanche in this sample is
\begin{equation}
W<W_1(L,d)=\exp[c_1 \ln^{1/3}(Ld)].
\label{Wnecessary}
\end{equation}
Under this condition, the thermal spot is guaranteed to be present in the sample and increases in size
up to the width $d$. After this, the avalanche dynamics is governed by the 1D law, as long as
\begin{equation}
2^{-(d+2x)d/2} e^{-x/\xi}\gtrsim 2^{-(d+2x)d},
\label{Q1D-avalanche}
\end{equation}
where $x$ is the linear size along the strip of the region hybridized with the ergodic spot to the right and to the left of the spot (hence $2x$ in the total number of ergodic states $N_s$).
Here, again, the factor $N_s^{-1/2}$ in the matrix element on the l.h.s. accounts for the full ergodicity.
For sufficiently weak disorder, $W \lesssim 2^{d}$,
the condition (\ref{Q1D-avalanche}) is satisfied for all $x$.
Since in the limit $L\to\infty$ a critical seed is always found in the sample, we conclude
that
\begin{equation}
W_c(L=\infty,d) \sim 2^{d}.
\label{WcdLinfty}
\end{equation}
We note that the condition $W\ll W_c(L=\infty,d)$ is compatible with the condition (\ref{2Dseed}) for the 2D geometry of the initial critical seed.

Comparing Eqs. (\ref{WcdLinfty}) and (\ref{Wnecessary}), we see that in large systems with
\begin{equation}
L>L_*(d)=\frac{1}{d} \exp\left[\left(\frac{d \ln 2}{c_1}\right)^3\right],
\label{Lstar}
\end{equation}
the condition $W<W_c(L=\infty,d)$ simultaneously implies that typically there is at least one seed in the system and the avalanche takes place.
This yields (see, however, below)
\begin{equation}
W_c(L,d) \sim 2^{d}, \quad L>L_*.
\label{WcLargeL}
\end{equation}
For shorter systems, $L<L_*$, the condition (\ref{Wnecessary}) is stronger than (\ref{WcdLinfty}) in restricting the strength of disorder from above for an avalanche to occur. Thus, for $L<L_*$ the bottleneck
for a full avalanche to occur is the presence of a critical seed in the sample, so that
\begin{equation}
W_c(L,d) \sim \exp[c_1 \ln^{1/3}(Ld)], \quad d<L<L_*.
\label{WcSmallL}
\end{equation}

Let us return to the case $L>L_*$. For large systems, one can find more than one critical seed in the whole sample for $W_c(\infty,d)<W<W_1(L,d)$. Each of the critical seeds would then grow up to the size
\begin{equation}
X_c\simeq d\frac{d \ln 2}{\ln W-d\ln 2}.
\label{Xc}
\end{equation}
This result generalizes Eq. (17) of Ref. \cite{DeRoeck2017a} to the quasi-1D setup.
If the number of critical seeds in the sample exceeds $L/X_c$, the whole sample will thermalize
by the overlapping finite avalanches even for $W>W_c(\infty,d)$.
The number of critical seeds is given by Eq. (\ref{PLdW}) when $P>1$.
The refined equation for $W_c$ is then given by
\begin{equation}
\frac{L}{d} \left(\frac{\ln W_c}{d\ln 2}-1\right)\sim L d\, e^{-c \ln^3 W_c}.
\label{refined}
\end{equation}
Clearly, for $d\gg 1$ the solution of Eq. (\ref{refined}) for $W_c$ yields a negligible
correction to the estimate (\ref{WcLargeL}).

Thus, the result for $W_c$ at $L>d\gg 1$ reads:
\begin{equation}
W_c(L,d) \sim \begin{cases} \exp[c_1 \ln^{1/3}(Ld)], &\quad d<L<L_*(d),\\
2^{d}, &\quad L>L_*(d),
\end{cases}
\end{equation}
where $L_*(d)$ is given by Eq. (\ref{Lstar}) and $c_1= c^{-1/3} \approx 1.57$ is the numerical coefficient
found from Eq. (\ref{cc}).

\section{Algorithm}
For our numerical simulations, we numerically integrate the TDVP equations using the scheme proposed in Ref.~\cite{Haegeman2016a}. We use the two-site implementation of the Open Source Matrix Product States library \cite{Wall2012a, Jaschke2018a} as well as a hybrid approach, implemented using the TeNPy library \cite{Tenpy} combining early evolution using the two-site algorithm with subsequent evolution using the one-site algorithm. In the case of the latter, we implement the Hamiltonian (1) from the main text, whereas in the former case we map this Hamiltonian to a representation with a synthetic dimension, with hopping and interactions represented by spin-orbit coupling as can be realized experimentally \cite{Celi2014a, Pagano2014a, Barbiero2019a}. In the geometric representation, the square lattice is mapped to a one-dimensional chain, which results in longer-range hopping terms that are implemented using matrix product operators \cite{Schollwock2011a}. We track the time evolution of various quantities, in particular the particle density at each site, the von Neumann entropy of entanglement and (for the two-site approach) the related entanglement spectrum. For more details on the algorithm and the differences between the one- and two-site approaches, we refer the reader to Refs.~\cite{Haegeman2016a, Paeckel2019a}.
\subsection{Two-site approach}
During time evolution, we demand that the error is bounded by requiring that the smallest value of the Schmidt numbers $\mathrm{min}(\lambda_i)$ in the entanglement spectrum \cite{Laflorencie2016a} $< 10^{-6}$ at any time step (cf.~Ref.~\cite{Doggen2019a}), using a maximum bond dimension $\chi$ of at least $128$ for stronger disorder and up to $512$ for the smallest systems and weakest disorder considered. During time evolution, we expand the bond dimension up to its maximum in order to keep the discarded weight $< 10^{-10}$ per site. In addition, we then check convergence with $\chi$. In practice, this procedure sets a minimum value of the disorder $W$ that we can consider, because the entanglement grows more strongly for weaker disorder. We use around $400$ independent realizations of disorder for each choice of parameters.
\subsection{Hybrid approach}
The hybrid approach is similar, but instead of considering the Schmidt values, we expand the bond dimension at each time step and switch to the one-site algorithm when the maximum is reached. We then again check for convergence with the maximum bond dimension $\chi$. The key difference between the two approaches is that the two-site approach leads to a truncation error that results from repeated singular value decompositions. This error is absent in the one-site algorithm, however, the two-site approach allows for dynamical control of the size of the variational manifold (controlled by the maximum bond dimension $\chi$). In practice, we switch from the two-site to the one-site approach at relatively short times $O(1)$ and use the two-site algorithm just to expand the variational manifold from the initial product state. The hybrid approach allows for studying larger systems, up to $10 \times 10$, for which we use $40$ realizations. We find that at strong disorder $\chi = 128$ is sufficient for $8 \times 8$-systems and $\chi = 192$ for $10 \times 10$.

\subsection{Comparison of the different schemes}
\begin{figure}
 \includegraphics[width=\columnwidth]{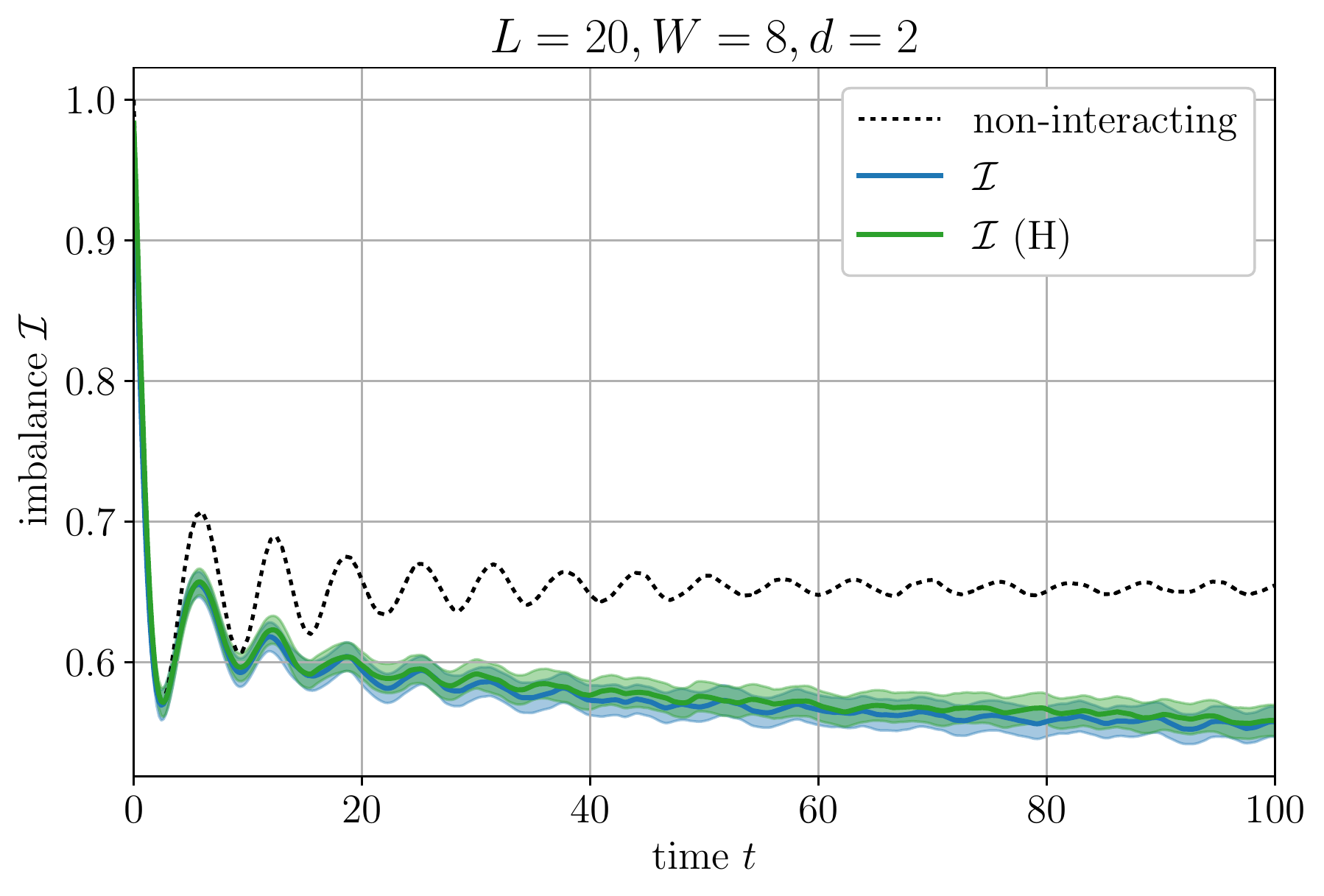}
 \caption{Averaged imbalance dynamics illustrating comparison between the two-site and hybrid (H) implementations with independent realizations of disorder, for $L = 20, W = 8, d = 2, \chi = 256$, and 400 realizations of disorder. The shaded region denotes a $2\sigma$-interval. The obtained power-law exponents are $\beta_\mathrm{two-site} = 0.039 \pm 0.006$ and $\beta_\mathrm{hybrid} = 0.037 \pm 0.006$ respectively.}
 \label{fig:comparison}
\end{figure}

In the main text, the different power laws obtained from the fitting procedure are compared between the two-site and hybrid algorithms. An explicit example of dynamics is shown in Fig.~\ref{fig:comparison}. The two curves are in good agreement for sufficiently strong disorder.

For weaker disorder, we observe better convergence for the hybrid approach. The truncation error of the two-site approach therefore clearly dominates the error in this regime. Hence, we confirm that also beyond the one-dimensional case \cite{Kloss2018a, Doggen2018a, Doggen2019a, Chanda2019a}, one-site TDVP (used in the latter part of the time evolution in the hybrid approach) is the superior MPS-based time evolution method. However, it remains not fully understood why the one-site TDVP algorithm seems to correctly capture long-time dynamics deeper into the ergodic regime \cite{Goto2019a}, where the two-site algorithm as well as comparable other approaches such as TEBD fail to converge with reasonable $\chi$. In addition to better convergence, we also observe faster computation times for the hybrid scheme. We attribute this to advantages of the geometric representation of the model, which allows for truncation (and hence computational speedup) in the transverse direction, while the synthetic representation captures the entanglement in the transverse dimension without approximations.

\subsection{Time step}

One of the benefits of the TDVP method is its stability with respect to the time step, which allows the use of a relatively large time step. In most cases $\delta t = 0.2$ provides converged results in step size, although we found that reducing the time step to $\delta t = 0.05$ is necessary for the largest systems and strongest disorder computed using the hybrid algorithm. This is attributed to the long-range terms that appear in this implementation, with hopping and interaction terms up to a distance $2d$.

\subsection{Convergence}

\begin{figure}
 \includegraphics[width=\columnwidth]{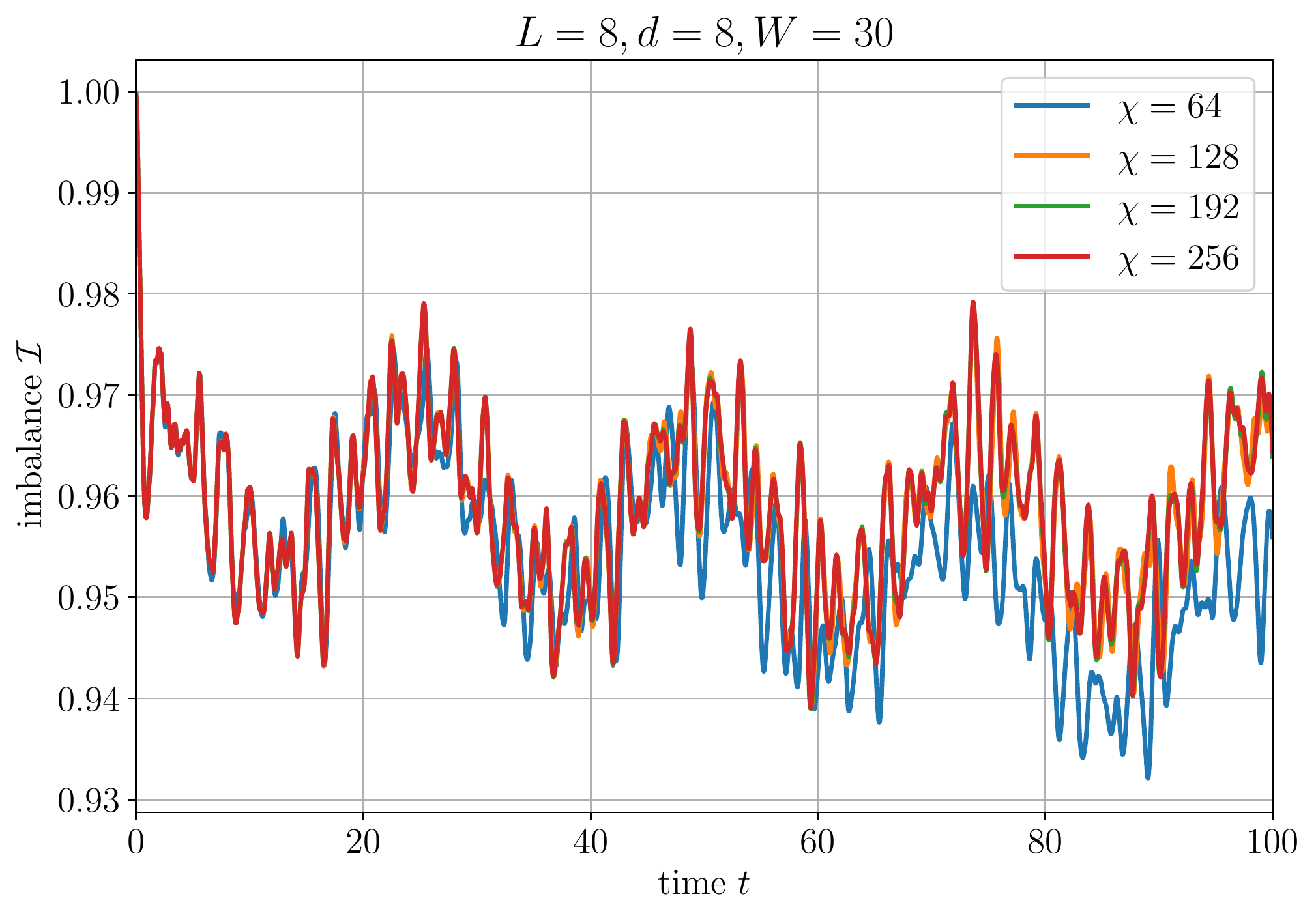}
 \label{fig:bench_8x8}
 \caption{Convergence of the simulations with respect to bond dimension $\chi$ for a single disorder realization, for $L = 8, d = 8, W = 30$.}
\end{figure}

Convergence of the algorithm relies on increasing the bond dimension $\chi$ until the results become independent of $\chi$. An illustrative example is shown in Fig.~\ref{fig:bench_8x8} where we display results for the same, randomly chosen, realization of disorder with different values of $\chi$. For the choice of parameters shown ($L=d=8$ and $W=30$), it is clear that $\chi = 128$ is sufficient for convergence. In this work, we always use at least $\chi = 128$. For the $10 \times 10$ system we use a larger value $\chi = 192$. For $\chi = 128$, $L=10$, $d=10$ and $W = 20$ we find $\beta = 0.038 \pm 0.002$, while for $\chi = 192$ we obtain $\beta = 0.039 \pm 0.004$. The different error estimates are due to a smaller number of disorder realizations used for $\chi = 192$ (160 instead of 400). We increase the bond dimension up to $\chi = 512$ for the quasi-1D case ($d \leq 4$) to reach deeper into the ergodic regime. Moreover, using $\chi = 256$ for small systems, $8 \times 2$ and $4 \times 4$, we can compute the dynamics exactly up to arbitrary times.

\begin{figure}
 \includegraphics[width=\columnwidth]{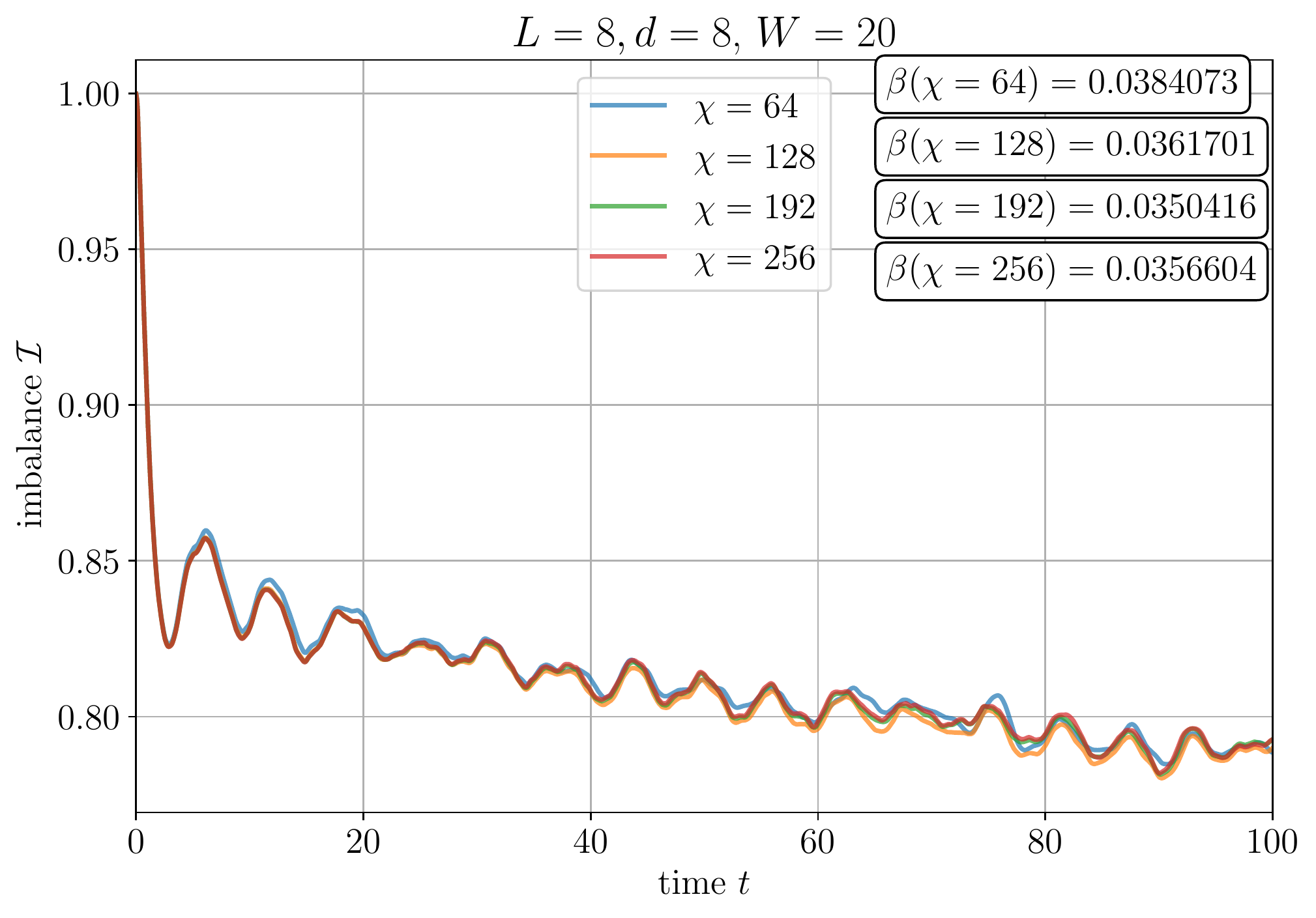}
 \label{fig:bench_8x8_W20}
 \caption{Convergence of the simulations with respect to bond dimension $\chi$ for 40 disorder realizations, for $L = 8, d = 8, W = 20$.}
\end{figure}

\begin{figure}
 \includegraphics[width=\columnwidth]{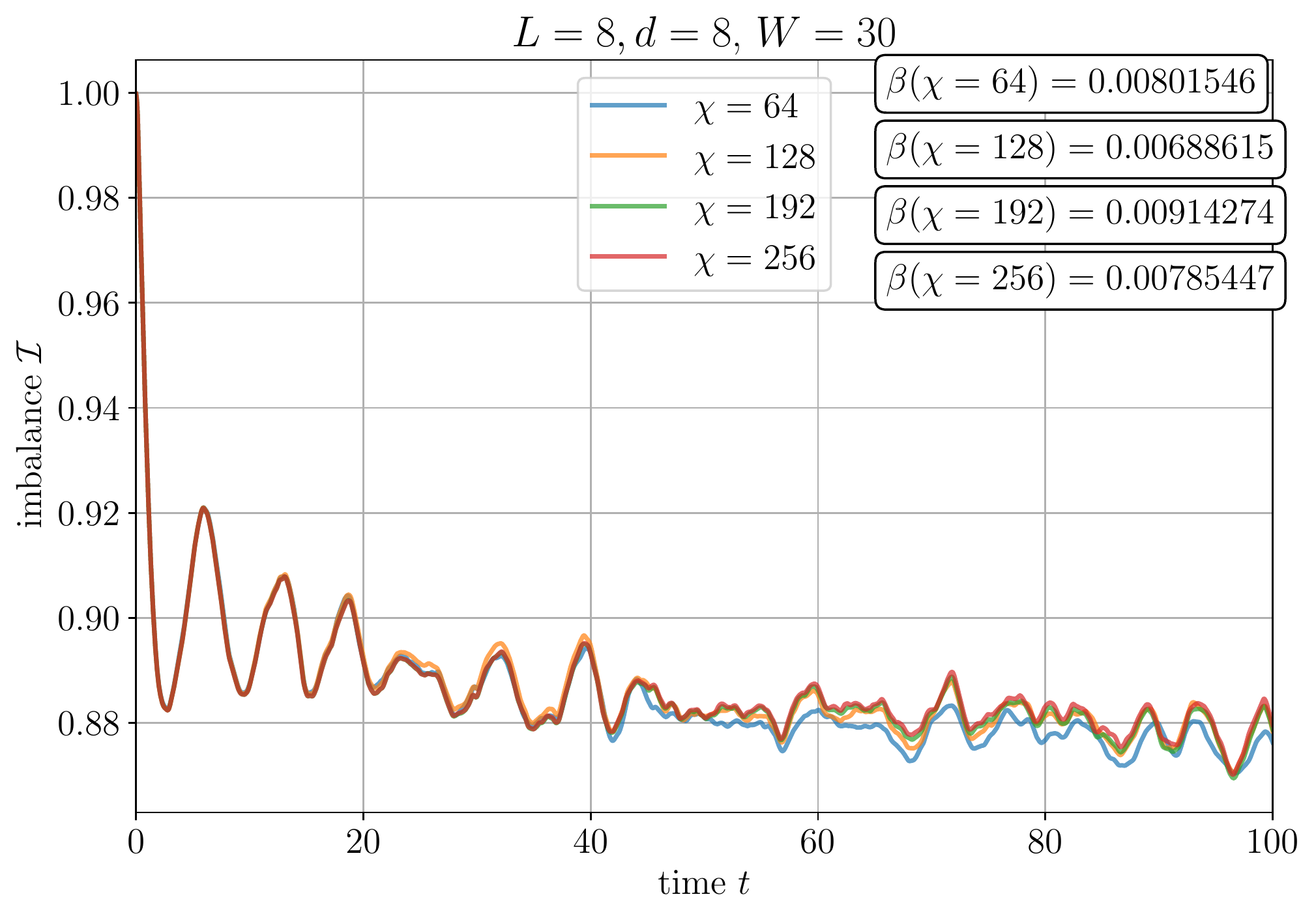}
 \label{fig:bench_8x8_W30}
 \caption{As Fig.~\ref{fig:bench_8x8_W20}, but for $W = 30$.}
\end{figure}

As a further illustration of convergence, we show in Fig.~\ref{fig:bench_8x8_W20}  the results for the imbalance averaged over the same 40 realizations for $L=d=8$, the disorder $W=20$ (well on the delocalized side), and various choices of bond dimension. Comparing the results  for $\chi = 128$ and $\chi = 256$,  we find an error in the predicted $\beta$ of around $0.0005$, smaller than the statistical error due to the finite number of realizations. This error decreases further with increasing $W$. (We recall that for our key results, such as the dependence of $W_c$ on the system size, only the behavior close to $W_c$ is important.)  Analogous results for $W=30$ are shown in Fig.~\ref{fig:bench_8x8_W30}.

\begin{figure}
    \includegraphics[width=\columnwidth]{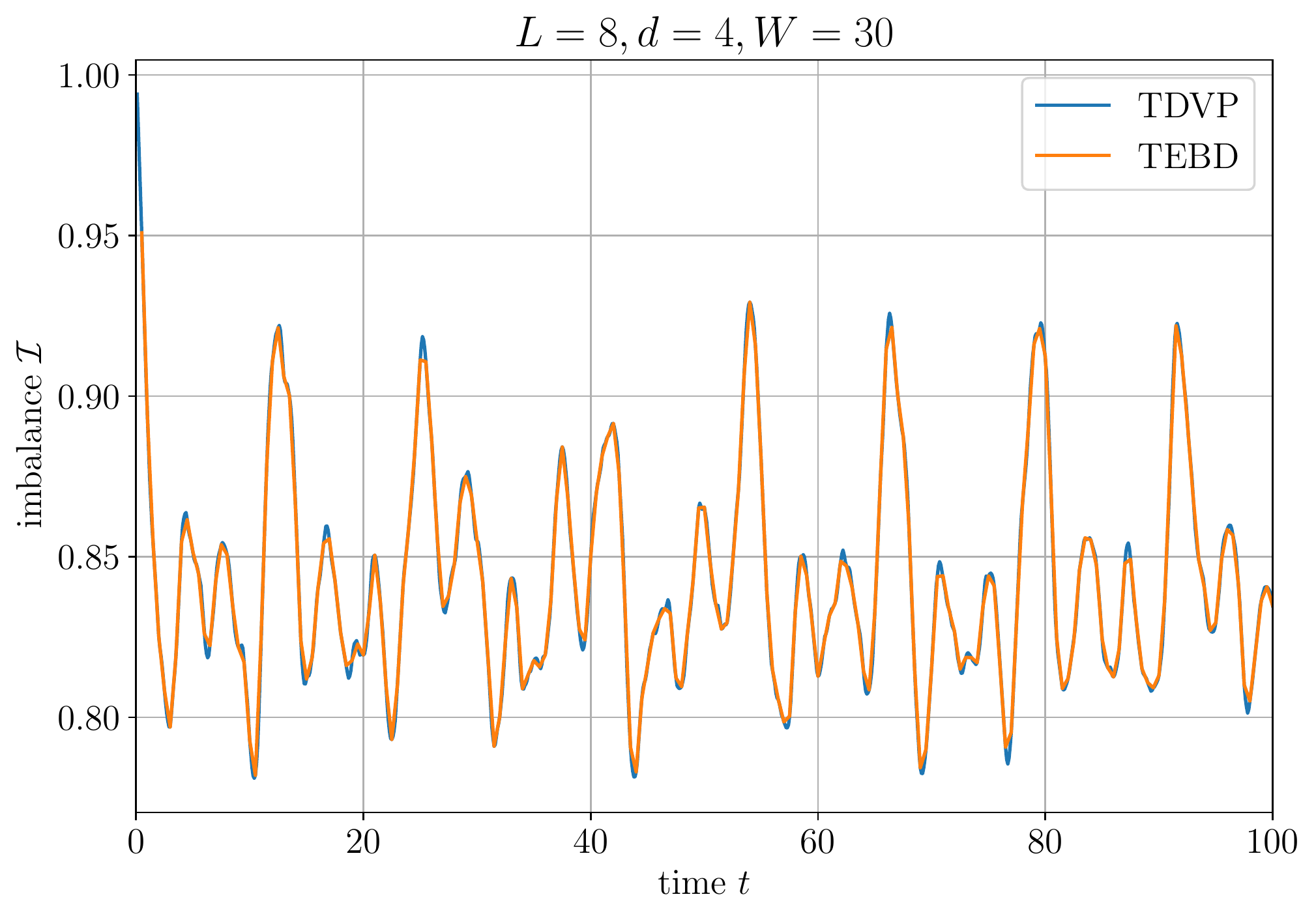}
    \caption{Comparison of TDVP to time-evolved block decimation (TEBD) for a single realization of disorder, for $L = 8, d = 4, W = 30, \chi = 128$.}
    \label{fig:tebd}
\end{figure}

Finally, we compare to other numerical methods. In the limit of strong disorder, various MPS methods perform similarly: they converge with relatively low bond dimension. On the ergodic side of the transition, however, TDVP tends to outperform other MPS-based approaches \cite{Doggen2019a,Chanda2019a} in terms of convergence of transport-related quantities (such as the imbalance). In Fig.~\ref{fig:tebd} we compare the TDVP and TEBD results for an $8 \times 4$-system, using the ``synthetic dimension'' (two-site) approach outlined above. The methods are in excellent agreement.


\subsection{Choice of the initial condition}

\begin{figure}
    \includegraphics[width=\columnwidth]{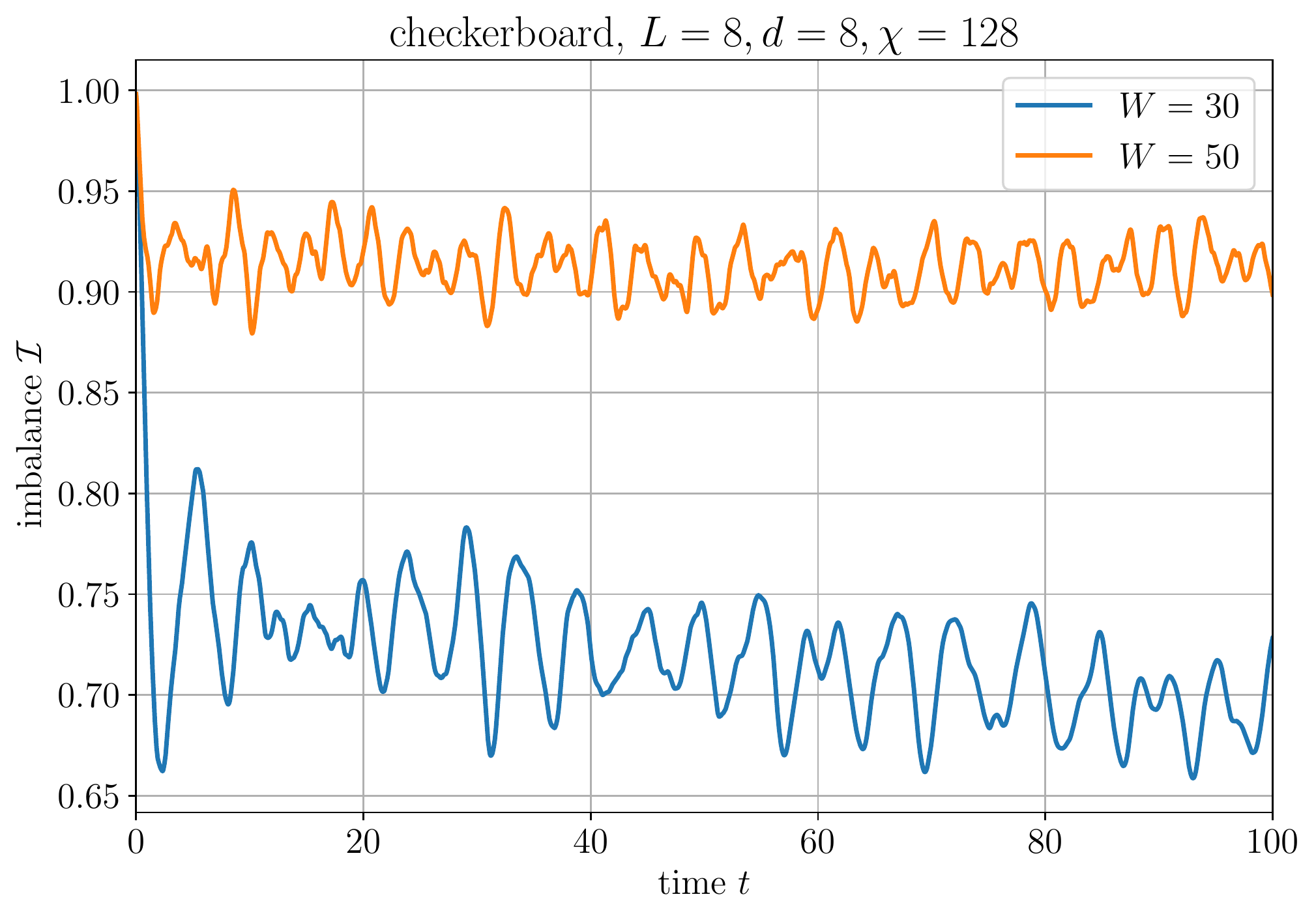}
    \caption{Imbalance dynamics for a single realization of disorder, for the ``checkerboard'' initial condition where the initially occupied site always neighbors an occupied site. For $W = 30$, the same disorder realization is used as in Fig.~\ref{fig:bench_8x8}.}
    \label{fig:my_label}
\end{figure}

The choice of initial condition should not affect the MBL transition. If the system is ergodic, then any initial state should thermalize; likewise, on the MBL side any initial state should remain localized. In our simulations so far, we have employed a striped initial condition, where ``rings'' are initially either fully occupied or unoccupied. In Fig.~\ref{fig:my_label}, we show, for a single realization of disorder, the dynamics of a ``checkerboard'' initial condition, where the initially occupied sites always neighbor only unoccupied sites. In this case, the imbalance is defined as the memory of the initial checkered state.

Comparing the results between the checkerboard and the striped initial conditions for $W=30$, there is a strong difference at short times, where the checkerboard imbalance drops several times more strongly than for the striped pattern. The decay at longer times weakens in both cases, with comparable power-law exponents of the decay. However, the amplitude of the change in the imbalance within the studied time interval after the initial drop is still much larger in Fig.~\ref{fig:my_label} than in Fig.~\ref{fig:bench_8x8}. These results can be explained as follows. In the short-time limit, the particles for the checkerboard initial condition have much more room to move around than the particles for the striped condition, which can only move in one direction close to $t=0$. After the initial falloff, dynamics is controlled by disorder, whereas the information about the initial condition shows up in the  difference in the definition of the correlators describing the imbalance in the two cases. Despite this difference, dynamics for the checkerboard initial condition freezes out at a comparable value of disorder $W$ between $W=30$ and $W=50$ (cf.\ Fig.~3 of the main text). Indeed, the data for $W=50$ in Fig.~\ref{fig:my_label} shows that the system is definitely on the localized side of the transition. We leave a quantitative analysis to future work.

\section{Definition of the critical disorder}

The critical disorder $W_c$ is estimated on the basis of saturation of the imbalance $\mathcal{I}$. However, we are restricted to the regime $t \leq 100$ in this work. Since we are interested in the late-time behavior of the system, we have to estimate the error $\delta\beta$ in the power-law exponent $\beta$ due to considering only finite times. Based on previous studies \cite{Luitz2016a, Doggen2018a, Chanda2019a}, we estimate the error in the range $\delta\beta \in [0.005, 0.01]$. This leads to the error bars in Fig.~3 in the main text, which are obtained according to the following procedure. We first fit the curve $f(W)$ to the data points for $\beta(W)$ close to $W_c$.  The error is then obtained by solving for $f(W_c) = \delta\beta$, with the ends of error bars corresponding to $\delta \beta = 0.01$ and $0.005$. 

\section{Fitting procedure}

For our determination of the power-law coefficient $\beta$, we use a standard Levenberg-Marquart non-linear fitting procedure, as implemented using the SciPy library's implementation of the appropriate MINPACK routine. The reported error is a $2\sigma$-interval determined using a statistical bootstrap procedure, where we consider 50 different samples of our data, with each realization sampled with $50\%$ probability. We find that this procedure yields error estimates that agree well with those obtained by including the error in the imbalance (i.e., twice the standard deviation of the average imbalance) in the fitting algorithm.

\begin{figure}
 \includegraphics[width=\columnwidth]{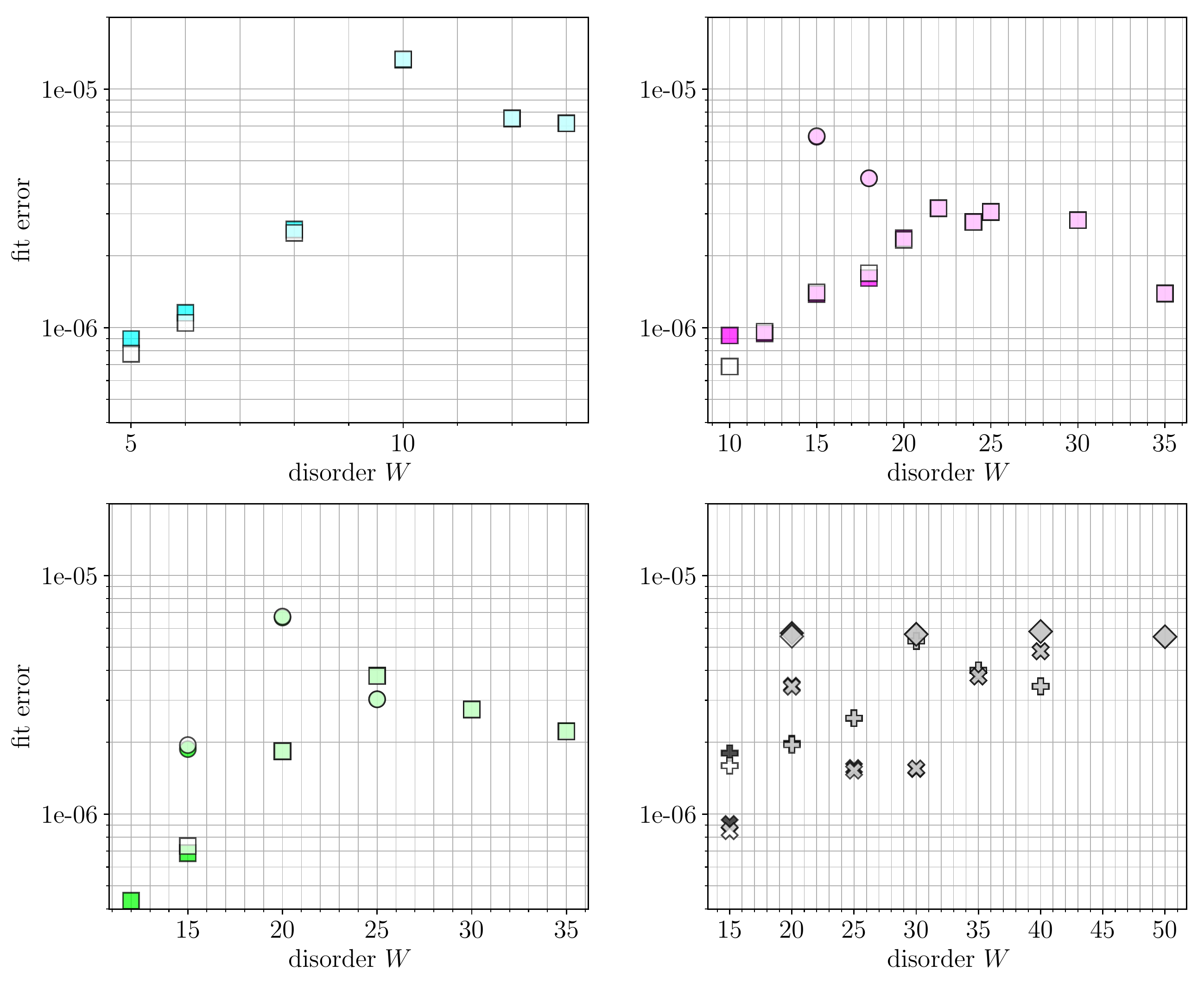}
 \label{fig:beta_errs}
 \caption{Fit error $\epsilon$ \eqref{eq:fiterror} as a function of disorder strength $W$ and system size $L, d$. The colors and symbols are identical to the main text, with an additional transparent white symbol for the fitting form \eqref{eq:2d_supmat}.}
\end{figure}

\section{Fitting form}

We have compared the power-law form $\mathcal{I}(t) \propto t^{-\beta}$ to the following form that has been predicted to result from Griffiths effects in 2D geometry
\cite{Gopalakrishnan2016a}:
\begin{equation}
 \mathcal{I}(t) \propto \exp\Big(-\gamma \ln^2 t\Big).
 \label{eq:2d_supmat}
\end{equation}
While in the $t \to \infty$ limit Eq.~(\ref{eq:2d_supmat}) decays faster than any power law, this requires exponentially long times for small $\gamma$. For realistic times, this formula provides a rather slow decay, very much similar to $t^{-\beta}$ with a small $\beta$.
We perform a fit in the window $t \in [50, 100]$ and compute the average $\chi^2$-error $\epsilon$:
\begin{equation}
 \epsilon = \frac{1}{N} \sum_{i=1}^N [f(t_i) - F(t_i)]^2, \label{eq:fiterror}
\end{equation}
where the number of data points is $N$, the fitting function is $f$ and the values of the data points are given by $F(t_i)$. The result is shown in Fig.~\ref{fig:beta_errs}. Clearly, for the considered time window the two forms are practically indistinguishable. The error of order $10^{-6}$ is likely dominated by disorder fluctuations.

\begin{figure}
 \includegraphics[width=\columnwidth]{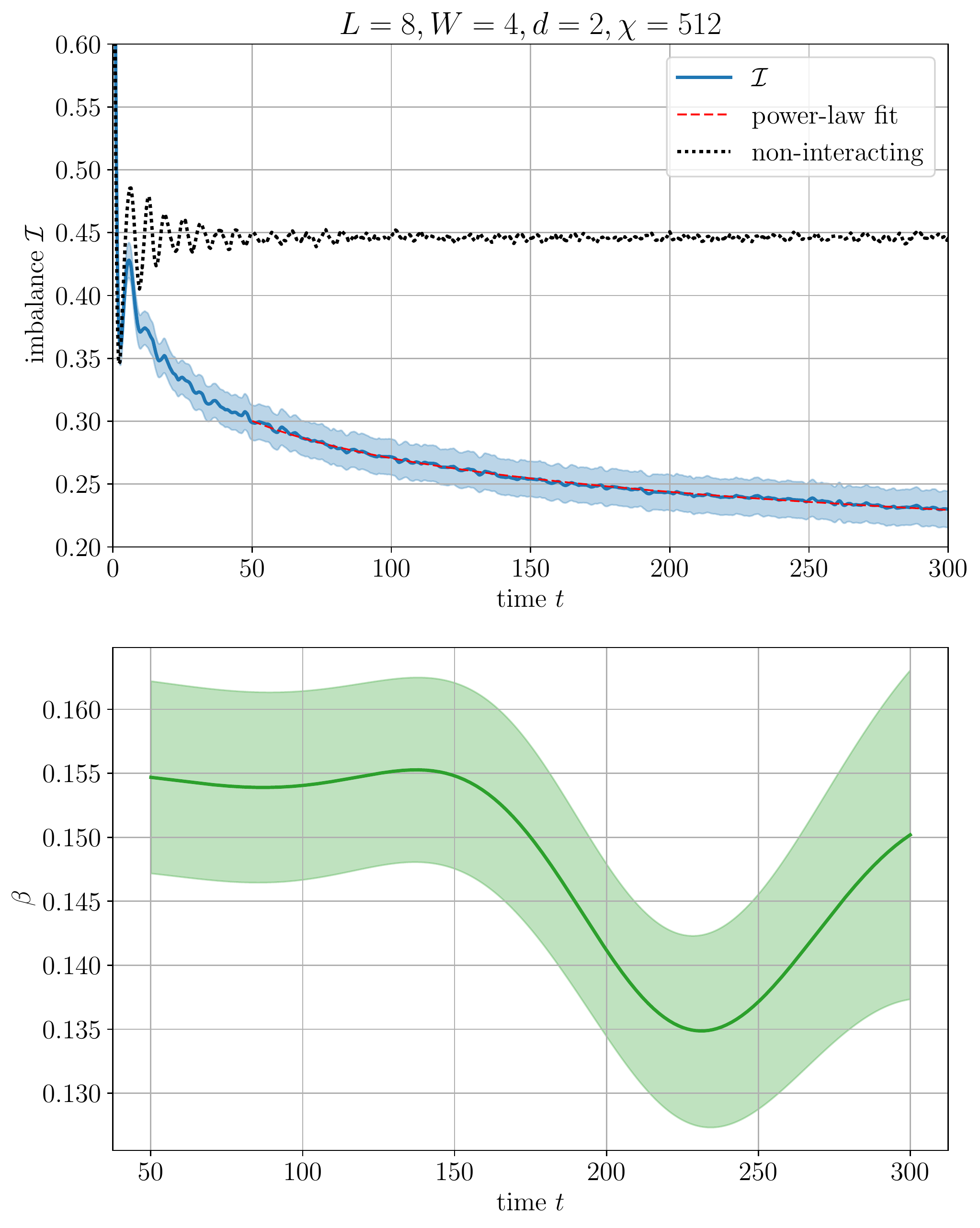}
 \label{fig:longtime}
 \caption{\textbf{Top}: averaged imbalance dynamics for a relatively small system with $L = 8, d = 2$ up to $t=300$ for disorder strength $W = 4$. The maximum bond dimension $\chi = 512$ is not reached during time evolution as $\chi = 256$ is sufficient for exact time evolution so that TDVP is equivalent to exact time evolution in this case. The power-law coefficient $\beta = 0.15 \pm 0.01$ as obtained from the window $t \in [50, 300]$ (compare to $\beta = 0.16 \pm 0.03$ obtained from the two-site approach in the window $t \in [50, 100]$). Results are computed using the hybrid TDVP method. The shaded region denotes a $2\sigma$-interval. \textbf{Bottom}: time-dependent power-law coefficient $\beta$, obtained by weighing data points according to a Gaussian with standard deviation $50$. The shaded region denotes a $1\sigma$-interval obtained by a statistical bootstrap, implying that $\beta$ is constant in time within $2\sigma$-error bars. }
\end{figure}

\section{Long-time dynamics  in quasi-1D geometry}

While entanglement growth limits the available time scales, this limitation is much less severe for smaller systems. An example is shown in Fig.~\ref{fig:longtime}. For a smaller system length, $L=8$, we proceed up to the time $t=300$. We observe no statistically significant deviations of the power-law behavior over the extended time window; for the window $t \in [50, 300]$ we obtain $\beta = 0.15 \pm 0.01$, while for the window $t \in [200, 300]$ we obtain $\beta = 0.14 \pm 0.02$. The pure power law form performs slightly better than the log-corrected form \eqref{eq:2d_supmat} (expected to be applicable to the 2D geometry), with about half of the $\chi^2$-error.

Alternatively, the quality of the fit can be assessed through a time-dependent fit coefficient $\beta(t)$ according to the procedure discussed in Ref.~\cite{Doggen2019a}, which weighs data points according to a Gaussian centered around the time $t$. The result is shown in the lower panel of Fig.~\ref{fig:longtime}.
While relatively small variations of $\beta(t)$ are  present, no clear trend that would show a systematic increase or decrease of $\beta(t)$ is visible.

We thus conclude that the imbalance decay remains to be well described by a power law also in an extended time interval, and the value of $\beta$ obtained from the fit over the larger time interval $t \in [50, 300]$ is the same as for $t \in [50, 100]$ within a few percent.

\section{Comparison of numerical results for $W_c$ with analytical predictions}

In this section, we provide a few additional comments concerning the comparison of the numerical results for $W_c$ shown in Fig. 3 of the main text with the analytical predictions of the avalanche theory, Eqs.~(3) and (4) of the main text. While Fig.~3 demonstrates a reasonable agreement between the numerical data and the analytical expectations, deviations are also evident. Here we comment on main sources of these deviations.

In this context, it is important to emphasize that our analytical expressions for the critical disorder are asymptotic formulas for large $L$.  In a finite system, there are corrections to scaling. For quasi-1D systems, the key requirement for the numerical determination of $W_c(\infty, d)$ [second line of Eq. (4)] is the condition $L \gg L_*(d)$. As the estimates in the main text show, it can be reasonably satisfied in numerical simulations (and in experiment) for $d \le 4$ but becomes essentially unrealistic for $d \ge 5$ because of the very fast growth of $L_*(d)$. The downward deviation of the $d=4$ point in the left panel of Fig. 3, as compared to the predicted exponential growth is in full agreement with this expectation: it can be attributed to the beginning of the crossover from the quasi-1D to 2D growth of $W_c$ [i.e., form second to first line of Eq. (4)]. 

In the 2D case, i.e.~for $W_c(L,L)$, the main source of deviations from the asymptotic formula [Eq. (3)] are finite-size corrections to scaling at small $L$. Such corrections that reduce $W_c$ at smaller $L$ are common for MBL systems. One source of corrections in the present case is that a thermal seed located at the boundary is less efficient at developing the avalanche instability. Analyzing all sources of finite-size corrections to scaling is a highly nontrivial task but, on general grounds, one may expect a relative correction to $W_c$ of the type  $(-1/L^\alpha)$ with $\alpha$ of order unity. This explains deviations at small $L$ in the right panel of Fig.~3; with increasing $L$ the numerical data approach the asymptotic curve, Eq. (3). 

\section{Comparison to previous numerical estimates of $W_c$ for quasi-1D geometry}

Let us compare our results for $W_c$ of quasi-1D systems with previous numerical works. The paper \cite{Wiater2018a} proposed estimates for $W_c$ based on level statistics obtained by ED of small systems. For the $d=2$ model, it was concluded on the basis of ED of systems of lengths from $L=6$ to $L=9$ that $W_c = 9.1 \pm 0.9$. These values of $L$ are much smaller than the values $L=20$ and $L=40$ for which we get saturation of $W_c$. Furthermore, inspection of original data in the inset of Fig. 4 of Ref.~\cite{Wiater2018a} shows that the crossing point there drifts systematically from $W \approx 8$ to $W \approx 10.5$ when the system sizes increases from $L=6$ to $L=9$. An analogous drift was observed earlier for $d=1$ models, as discussed above. Thus, the value $W_c = 9.1 \pm 0.9$ obtained in Ref.~\cite{Wiater2018a} by a fit to the data for small systems underestimates the large-$L$ limit of $W_c$, in consistency with our result $W_c(\infty,2) \approx 12$.  Similarly, for $d=3$ Ref.~\cite{Wiater2018a} used systems of lengths from $L=3$ to $L=6$ and concludes that $W_c = 12.1 \pm 1.6$. Again, inspecting Fig. 4 of Ref.~\cite{Wiater2018a}, we observe a strong drift of the crossing point: from $W \approx 10$ to $W \approx 14.5$ in this interval of lengths. Thus, there is no contradiction between these values and our results obtained for considerably larger systems, $W_c(8,3) \approx 20$ and $W_c(20,3) \approx 26$.
The value $W_c = 12.1 \pm 1.6$ of Ref.~\cite{Wiater2018a} is thus a  characteristic of small-$L$ systems and is (at least) twice smaller than the large-$L$ limit $W_c(\infty,3)$.
In Ref.~\cite{Baygan2015a}, a two-leg spin ladder was studied that is slightly different from the model (1) but is closely related to it. This paper also used ED of small systems, and the above discussion fully applies in this case as well: the value $W_c = 8.5 \pm 0.5$ stated in Ref.~\cite{Baygan2015a} substantially underestimates the $L \to \infty$ critical disorder of the two-leg ladder.

In Ref.~\cite{Hauschild2016a}, domain-wall melting was considered as a probe of MBL (as an alternative to the imbalance decay), following the experiment \cite{Choi2016a}. If one would be able to proceed controllably to arbitrary long times, both approaches would give the same $W_c$. It turns out, however, that the dynamics of domain wall spreading is much slower. Indeed, a dramatic difference can be observed already in the non-interacting case: for the domain wall initial condition of Ref.~\cite{Hauschild2016a}, the authors report requiring $t > 1000$ to access the asymptotic regime, whereas for the charge-density-wave initial condition used here we observe saturation of the imbalance to its asymptotic value (plus oscillations which average to zero) at much shorter times $t \sim 10$ (see the black dotted line in Fig.~1 of the main text). In the interacting case, the numerically accessible times are limited, which makes it advantageous to use the imbalance. While the authors of Ref.~\cite{Hauschild2016a} estimate the critical disorder as $8 \lesssim W_c \lesssim 10$ for a long $d=2$ system, it is seen in their Fig. 4 that the domain wall dynamics has not really saturated either for $W=8$ or for $W=10$. In this sense, results of Ref.~\cite{Hauschild2016a} compare well to our findings for these values of $W$ but statistical uncertainties in Ref.~\cite{Hauschild2016a} are higher. We thus conclude that the domain-wall dynamics of Ref.~\cite{Hauschild2016a} is not in contradiction with our result for the critical disorder, $W_c(\infty,2) \approx 13$.

\bibliography{ref}